\RequirePackage{fix-cm}
\documentclass[smallextended]{svjour3}       % onecolumn (second format)
\smartqed  % flush right qed marks, e.g. at end of proof
\usepackage{graphicx,url}

\newcommand{\ket}[1]{\left| #1 \right>} % for Dirac bras
\newcommand{\bra}[1]{\left< #1 \right|} % for Dirac kets
\begin{document}

\title{Elements of reality in quantum mechanics%\thanks{Grants or other notes
}
\subtitle{}


\author{Geoff Beck.
}


\institute{G. Beck \at
              School of Physics, University of the Witwatersrand - Private Bag 3, WITS-2050, Johannesburg, South Africa
              \email{geoffrey.beck@wits.ac.za}           %  \\
}

\date{Received: date / Accepted: date}

\maketitle

\begin{abstract}
The notion of the Einstein-Podolsky-Rosen (EPR) ``element of reality" is much discussed in the literature on the foundations of quantum mechanics. Recently, it has become particularly relevant due to a proposed criterion of the physical reality of a given quantum mechanical observable [A. L. O. Bilobran and R. M. Angelo, Europhys. Lett. \textbf{112}, 40005 (2015)]. We examine this proposal and its consequently related measure of non-locality [V. S. Gomez and R. M. Angelo, Phys. Rev. A \textbf{97}, 012123, (2018)] and argue that the criterion is ill-described as quantifying physical reality without introducing serious inconsistency with the basic notions of realism that under-gird enquiry. We agree that this reality criterion demonstrates, along with the famous GHZ results, that general quantum observable values make for poor elements of reality. However, we also argue that this does not mean no such elements of reality are to be found in quantum theory. By arguing for, and adopting, probability distributions as these elements of reality instead, we demonstrate that the criterion of physical reality is actually one of observable predictability. We then examine the relationship of realism-based non-locality to the Bell form and find that, despite the flawed premise, this measure does indeed codify non-locality that is not captured by Bell inequalities.
{Quantum foundations}% PACS, the Physics and Astronomy
{Entanglement and quantum non-locality}
\end{abstract}

\section{Introduction}
\label{intro}

The discussion of ``elements of reality" may seem like pure philosophy without any operational relevance. However, it is much discussed in recent physics literature on quantum theory (see  \cite{psi1,psi2,psi3,psi4,psi5,psi6,psi7} for prominent examples). Additionally, \cite{bilobran2014} proffers a quantitative ``criterion of the physical reality of a quantum observable". The measure put forth is heavily couched in the language of quantifying whether observables are ``elements of reality" in the sense invoked by the EPR paper~\cite{epr}. This is further used in \cite{gomes2018} to produce a measure of non-locality that is sourced from how much the reality of quantum observables is affected by distant measurements. This is of some significance, as it is known that quantum systems can exhibit non-locality that is not captured by the Bell inequalities~\cite{nonlocal1,nonlocal2,nonlocal3} or similar measures of this phenomenon. Thus, it is important that the premises of these measures of reality and non-locality are interrogated to determine whether they actually quantify what they claim to.

The aforementioned criterion of reality asks the operational question: ``how definite was the value of the observable $A$ given a state preparation $\rho$?" If $A$ had a pre-determined value then it has reality $1$, if it has spectrum of values with assigned probabilities then it has reality $< 1$. This therefore assigns the designation ``element of reality" to any value of a quantum observable that can be predicted with certainty, particularly this applies to actual measured values. Intuitively this seems reasonable as quantifier of reality, as perhaps the most basic commitment of scientific realism is the idea that there is a pre-existing world that can be interrogated by our measurements\footnote{\url{https://plato.stanford.edu/entries/scientific-realism/}}. Therefore, if some quantity is not pre-defined prior to measurement then it is not completely real. This agrees with the EPR definition as well, as ``elements of reality" are physical quantities that can be predicted with certainty with disturbing the system in question. Thus, the measure from \cite{bilobran2014} is claimed to extend a reality criterion to mixed states as well eigenstates of observables. Following this line of reasoning we see that a general quantum state may have no real properties at all until it is measured and some observables obtain real values.
This follows in the foot steps of the famous GHZ scenario~\cite{ghz,ghz-exp}, in which it is shown that measurement of a system of three entangled qubits is inconsistent with the notion that all of spin values for each axis and qubit were pre-assigned by state preparation. We can see the commonality with \cite{bilobran2014} in that real properties emerge from interaction with quantum systems (like measurement). In the GHZ case this can then be construed as a refutation of the idea of the EPR~\cite{epr} ``element of reality"~\cite{mermin1990}. Despite their use of the EPR criterion it is hard to avoid the same argument for \cite{bilobran2014}, as we are similarly lead to conclude that for any general scenario real properties emerge from the interaction of quantum systems rather than being pre-existent. This tension with the EPR criterion does far more than these arguments want to deliver, as it is a tension with the notion of some underlying pre-existent reality, which undermines the very basis of scientific enquiry. Stated as baldly as this, the over-reach of claiming that some reality quantification demonstrates that real properties can only emerge from interaction with quantum systems and that therefore the state corresponds to no ``elements of reality" itself becomes almost paradoxical. As one must immediately ask why the real properties that emerge from interaction with a given state have statistical predictability if the state itself possesses no real properties until measured? It seems then that by showing that values of observables do not generally correspond to elements of reality it is concluded that the EPR definition has failed, but the persistence of the predictability needed by the EPR criterion should give us pause. Perhaps, if a quantification of reality is inconsistent with some basic realism, that must indeed premise its quantification, then it was not quantifying what we believed it to be.

Thus, the premise of this article is to study the notion of the physical reality of quantum observables as presented by \cite{bilobran2014} and to show that by assuming elements of reality are assigned to measured observable values it only succeeds in quantifying the predictability of those observables. In addition to this it will be shown that the GHZ result does not provide the grounds for a wholesale rejection of EPR ``elements of reality" in quantum mechanics, contrary to the argument in \cite{mermin1990}. This argument is built upon the notion that all of these claims about reality are actually claims about classicality, and that the conflation of these two explains the inconsistency with scientific realism outlined above. It is argued that the very fact we can predict the GHZ outcome with certainty immediately undermines the claims of a failure of the EPR criterion. The probabilities of measurement outcomes in quantum mechanics are then put forward and demonstrated to posses all the necessary properties of an EPR element of reality. The failure of observable values as general EPR  elements of reality stems from the elementary observation that only statistical objects could fulfil this role in quantum theory as this explicitly statistical formalism cannot be used to infer the ontological status of non-statistical objects~\cite{popper1982}. This serves to undermine any claims about quantifying the reality of observables, reducing the conclusions of \cite{bilobran2014} to showing that observable values' ontological status is outside of the scope of the quantum formalism and what they quantify is purely observable predictability. The failure of the central premise of the non-locality criterion of \cite{gomes2018} requires us to examine what it is actually measuring. Despite the shaky foundation, it is shown that it does indeed quantify non-locality that it not captured by the Bell inequality as claimed by the authors of \cite{gomes2018}. However, the relationship discussed in \cite{info-reality} is reduced to a trivial one of a complementarity between predictability and information.

The argument is presented as follows: in section~\ref{sec:reality} we discuss the criteria presented in \cite{bilobran2014,gomes2018} and examine some simple unintuitive consequences that they represent. In section~\ref{sec:properties} we discuss what properties of quantum systems can actually have their ontological status inferred from empirical measurements and argue that the EPR element of reality is alive and kicking in section~\ref{sec:eor}. We then proceed to examine realism-based non-locality in section~\ref{sec:local}. Our conclusions are summarised in section~\ref{sec:conc}. 

\section{\label{sec:reality}The reality of observables}
The measure of the physical reality of an observable presented by \cite{bilobran2014} is criterion based upon whether the value of a given observable was predefined by the preparation of the state $\rho$. To be more concrete they imagine a situation where ideal state tomography is being carried out on a prepared stated $\rho$. However, before the state can be measured it is intercepted and has an observable $A$ measured without the result being revealed. Thus, the state being measured is in fact
\begin{equation}
\phi_A(\rho) = \sum_a \ket{a}\bra{a} \rho \ket{a}\bra{a} \; ,
\end{equation} 
where the sum runs over eigenvalues $a$.

The ``irreality" of $A$ given $\rho$ is then defined as
\begin{equation}
\mathcal{I}(A\vert\rho) := S(\phi_A(\rho)) - S(\rho) \; ,
\end{equation}
with $S$ being the von Neumann entropy. This quantity is non-negative and vanishes only when $\rho = \phi_A(\rho)$. Thus, it will always vanish if the state $\rho$ is an eigenstate of $A$. In agreement with the EPR criterion, eigenstate preparations are always elements of reality, but this also attempts to generalise beyond this and quantify the reality of mixed states as well. This criterion is premised, as is apparent, on the idea that observable values become real once they have been measured. 

It is immediately apparent that $\mathcal{I}$ will be $< 1$ for many choices of mixed state $\rho$. Thus, for a general quantum state, observables like $A$ will not constitute elements of reality (in agreement with the GHZ results~\cite{ghz}). We might then be tempted to compare the reality of quantum versus classical physics, with the perplexing result that classical observables are elements of reality but quantum ones are not in general. This suggests that despite being ostensibly more fundamental than classical physics, and thus more ontologically real, quantum mechanics references no general elements of reality. This is inconsistent with any basic notion of scientific realism, and with the idea that classical physics is a derivative reality sourced from something more fundamental. Alternatively, we could argue that the notion of ``elements of reality" is actually invalid, and that reality results from the interaction of quantum systems. However, although this sounds more sophisticated than the preceding argument these both seem to be easily restated with ``classical" substituted for ``real" without losing anything in the process. This suggests that we should be cautious of the use of $\mathcal{I}$ as irreality, as this may well be misleading. However, $\mathcal{I}$ does obviously characterise the degree to which we can predict the value of an observable $A$.

Of additional importance are the cases in \cite{bilobran2014}, where it is shown that observables for non-separable systems are not maximally real and that this is also true for simultaneous measurement of non-commuting observables. These cases are instructive as they suggest that $\mathcal{I}$ may in fact be a measure of the classicality of an observable. As the principle cases in which irreality is to be found are those where the quantum and classical mathematical formalisms diverge. This is reinforced when we note that in classical physics both theoretical predictions of statistics and individual observable values can be compared to empirical results and thus can be tested for some ontological value. However, in the quantum case this option is not available. This is because all quantum predictions stem from an inherently statistical formalism and it is elementary that no statistical premise may lead to conclusions about non-statistical properties (this will be elaborated on in the next section). This simple line of argument immediately suggests to us that quantum mechanics cannot operationally quantify the reality of a given observable, only the reality of associated statistics. So in asking a quantification of the predictability of an observable to act as a measure of reality we are inherently demanding that reality behave classically. 

In order to examine this issue in more detail we must proceed to study what properties of a quantum system can actually have their ontological value determined. That is, we must ask the question of what properties of quantum systems will the success or failure of experimental tests reflect upon. This question is necessary as these are the only quantities that can reasonably satisfy the EPR definition and qualify as elements of reality.

\section{\label{sec:properties}Properties of quantum systems}
%Elaborate on what constitute properties of quantum systems. Using pressure as a comparative classical quantity (both for GHZ as it is one-shot and for relying on underlying stats not x's). 
Quantum mechanical observables do not constitute well behaved ``elements of reality", as evidenced both above and by the GHZ experiment. Thus, the notion of an ``element of reality" should be disposed of, quantum properties are not pre-determined and result from interaction with the system itself. This argument seems to be robust, but there is a crucial twist that we need to follow, we implicitly assumed that only the values of the observables themselves can constitute the properties of quantum system. Without this assumption it is no longer clear that we are forced to reject the existence of ``elements of reality", perhaps it is simply that values of quantum observables do not constitute these elements within the evidently statistical framework of quantum mechanics? This argument should especially be suggested to us by the fact that the GHZ outcome is predictable with certainty, surely something real must under-gird this happen-stance? Clearly the EPR definition has not been truly disposed of. We should expect this, as the criterion is analytic~\cite{maudlin_bell}, and to declare it does not apply at all in quantum mechanics must also throw out the root notion of scientific realism: that there is some independent reality to interrogate. What has happened then is that the ``element of reality" that allows the GHZ result to work at all, rather than output subjective gibberish, is not the individual value of observables within quantum theory. 

This problem should be expected, as there is, in general, no measurement that can be made on a quantum state $\rho$ whose correspondence with prediction would depend upon the ontic status of the value of an observable $A$. In other words, quantum mechanics predicts statistical patterns to the behaviour of quantum states and thus by comparing observed and predicted outcomes of experiments we determine whether outcome probabilities $p(a|A,\rho)$ have any basis in what is real. If these probabilities had no correspondence to an underlying reality then this would be revealed by their failure to match the observed statistics. Furthermore, probabilities obviously cannot be tested by singular statements of measurement. Thus, within the framework of quantum mechanics there is no singular measurement of an observable $A$ that depends upon the correspondence of the value of $A$ with some underlying reality. For this to occur, we must be able to derive non-statistical conclusions from statistical premises~\cite{popper1982}. Simply put, the theoretical formulation of quantum mechanics makes it clear that the definiteness of values of observables $A$ cannot be consistently used as criterion of what has ontological value within the quantum paradigm. This does not necessarily imply values of $A$ are irreal, merely that there is no such information within the quantum formalism. This is then what we learn from both the work of \cite{bilobran2014} and GHZ, rather than grand claims about the nature of reality.

The eagle-eyed reader has already spotted that this argument seems to fail for GHZ cases, as their central importance is that they are ``one-shot" measurements that do not require us to accumulate statistics. However, the notion of the ``one-shot" measurement needs some careful attention. To elucidate it consider the classical example of the pressure of a gas which we can predict through statistical physics. This is a ``one-shot" measurement, but the success of our predictions reflects upon our statistical assumptions about velocity distributions of particles in the gas rather than the positions and velocities assigned to any given particle. Even though, barring error analysis, we did not need to accrue statistics, we did not avoid the fact that all we were testing was the validity of our statistical assertions. The same is true for the GHZ case, as the outcome prediction is a statistical relationship between three individual observables. There is only one predicted pattern, but, there are many combinations of individual observable values that would satisfy this prediction and these actual individual value assignments are not themselves predicted. Additionally, we must note that we did not step outside the states and amplitudes framework of quantum mechanics and use a separate apparatus to produce results that were premised on the validity of individual observable values. Thus, we are still implicitly testing our statistical assertions about the behaviour of quantum states, regardless of whether the measurement required us to accrue statistics or not. We can appreciate this by considering that our prediction of $p(x)$ cannot be used to predict the an individual trajectory in $x$, in GHZ this manifests as the success of the GHZ result does not allow us to infer the value of any one of three observed spins on its own, we can at best assign it the 50/50 chance of being up or down. 

\section{\label{sec:eor}Elements of reality reborn}

We have seen in the previous sections that observables are not generally adequate as elements of reality in quantum mechanics. This does not mean, however, that such elements are not to be found. An obvious quantity that is predictable with certainty without disturbing the state of a system is provided in the form of the probability distributions for outcomes of the measurement of a given observable $A$, $p(a|A,\rho)$ (which relates closely to the claims in \cite{psi1,psi2,psi3,psi4,psi5,psi6,psi7} of the quantum state encapsulating all relevant elements of reality). A reader might point out that the EPR criterion clearly specifies a ``physical quantity" and thus might question whether $p(a|A,\rho)$ could satisfy this requirement. To clarify this we can consider an illustrative classical example, that of a Maxwell-Boltzmann velocity distribution. Such a distribution can be used to predict the properties of an approximately ideal gas. Were the distribution not a statement of the physical properties of the gas itself this would not be possible without some violation of physical law or other miracle. Thus, it is argued that objects like $p(a|A,\rho)$ could not be used to predict outcome frequencies, and thus observable averages, unless they were reflecting physical properties of the system represented by the state $\rho$.

The argument in favour of the ontic nature of probability distributions still feels weak. All that was required is to satisfy a no-miracles argument, and $p(x)$ may still depend strongly on how much information a given observer has. However, consider the Maxwell-Boltzmann case again, this time we imagine that we have precise knowledge of the position and momenta of all particles at some time $t$ and perform $N$-body simulations to determine the statistical properties of the gas by averaging. If the statistical properties predicted by the Maxwell-Boltzmann distribution match those measured, the only way this could be true would be if our $N$-body simulation yielded the same answers. This of course implies that probability distributions predicted by our statistical theory are informationally complete (within their domain of testability) and thus have no meaningful observer dependence. This can be viewed from another angle by considering the operational aspects of the maximum entropy method advocated by Jaynes~\cite{jaynes2003}. In this case the accurate modelling of statistical properties requires that we have an exhaustive list of constraints on our system. For the modelling of a physical system these constraints can only take the form of assertions about relationships between physical properties. In other words if our predictions for $p(x)$ fail to fail we have strong evidence that our probability distributions are informationally complete (on a statistical level of course), as we neither have too many nor too few constraints. In quantum mechanics we have no analogous $N$-body case to appeal to, however, we do have the generality of the above argument and the theorems asserting that the wavefunction $\psi$ is informationally complete within the statistical form of quantum mechanics (for instance \cite{psi1,psi2}). The consequence of this is that we can be safe in assuming our $p(x)$ to be ontic rather than epistemic, this conclusion is also reinforced by the failure of the EPR argument for the incompleteness of quantum mechanics. 

Significantly the quantities $p(a,b|A,B,\rho)$ are mutually definable with the same certainty regardless of whether or not $A$ and $B$ commute. More importantly, their representation of physical properties is easily inferred from their predictive success, unlike individual measurements of $A$ or $B$, as argued above.

A reader might also be unsatisfied by the fact that these probabilities are not the properties of an individual electron for instance. But rather, the properties of an electron within a given situation. Once again we stress the analogy to the statistical properties of a gas, pressure for instance requires the presence of a confining force but is still determined by the internal physical properties of the gas itself, despite the additional dependence on the external situation. This is no different to the observation that a classical particle behaves differently in different potentials, but, knowledge of these potentials can be used to deduce the internal properties (mass and charge) of the particle itself from this behaviour. These internal propensities to respond to external influence are also what we extract in analogous quantum situation. They are merely all the information available within the structure of quantum theory, as this structure codifies what we can seek to measure when testing the theory.

The choice of $p(a|A,\rho)$ as the elements of reality is important, as it does not lead us to conclude quantum mechanics is less real than the classical case or require us to conflate real with classical. This is because the statistics inferred from quantum mechanics will have more predictive success and thus can be seen as unambiguously more real. This clearly establishes that neither a criterion for the reality of observables, or the abandonment of ``elements of reality" establish a consistent/workable idea of realism in the same manner as adopting $p(a|A,\rho)$ (and perhaps $\rho$ itself) as the elements of reality. Moreover, it clearly demonstrates that conflating the predictability of observables with their reality, without a careful examination of the theoretical formalism, leads to deep inconsistencies. Thus, the quantification from \cite{bilobran2014} is not a criterion of reality at all, we are forced to stop at ``criterion of predictability".

\section{\label{sec:local}Realism-based non-locality}

Having established what \cite{bilobran2014} actually quantify, we must now turn to the extension of this work to quantifying non-locality that stems from quantum measurements influencing the reality of observables. This is the realism-based non-locality of \cite{gomes2018}. In short, if we envisage a scenario with two entangled sub-systems each prepared in state $\rho$ and sent to be measured in a space-like separated arrangement. An additional system is prepared in a known state and allowed to interact locally with only one of the sub-systems prior to measurement, with the new state now being $\Phi_B(\rho)$ in this subsystem. The non-locality in question is then whether the reality of an observable $A$ has been changed in the sub-system that did not suffer the local perturbation. This is quantified by
\begin{equation}
\Delta \mathcal{I} (A,B|\rho) = \mathcal{I}(A|\rho) - \mathcal{I}(A|\Phi_B(\rho))
\end{equation}

In the preceding section with established that $\mathcal{I}$ was in fact a measure of predictability of observables, rather than their reality. So we are now equipped to see what $\Delta\mathcal{I}(A,B\vert\rho)$ actually tested. This is evidently the effect of distant measurements on the predictability of measurement outcomes, or whether $p(a\vert A, B,\rho) = p(a\vert A,\rho)$ under these conditions. This provides a more relaxed test for the effects of non-separability than a Bell inequality, which demands a correlation in $p(a)$ and $p(b)$ that exceeds what is possible under a hypothesis of local causality. The quantity $\Delta \mathcal{I}$ instead informs us of \textbf{any} non-separability related correlation between the outcome probabilities, as for separable states $\Delta \mathcal{I} \to 0$ so separable correlations will perforce be ignored. This means that, despite the flawed premise of the reality criterion, the non-locality measure does indeed capture aspects of non-separability that are not captured by the Bell inequality as claimed by the authors in \cite{gomes2018}.

This has consequences for other work based on \cite{bilobran2014}, such as \cite{info-reality} where a complementarity between information and the reality measure of observables is discussed. With the arguments presented here we can see that this relationship is reduced to triviality, as it is evident that there is a complementarity between Shannon-like information~\cite{shannon1948} measures and predictability of the observables encoding this information.

\section{\label{sec:conc}Conclusion}
In this work we examined whether cases which quantify the reality of observables in quantum mechanics~\cite{ghz,bilobran2014} actually reflect in some way upon the underlying ontological situation. It shown that, because quantum mechanical predictions are inherently based upon statistical assumptions, that only statistics can have their ontological status inferred from the success or failure of quantum mechanical predictions without introducing contradictions with basic tenets of scientific realism. This means that \cite{bilobran2014} does not in fact quantify the reality of observables but merely their predictability (which is also more easily defined operationally). However, despite the failure of the underlying premise, the non-locality measure proposed in \cite{gomes2018}, based on \cite{bilobran2014}, does indeed quantify non-locality not captured by Bell inequalities. This non-locality, however, is not related to the reality of observables at all, merely how their predictability varies as a result of non-separability.  

In addition to this, we refute arguments that we should discard the idea of elements of reality in quamtum mechanics, or that no such elements can exist, by demonstrating that probabilities assigned to various outcomes satisfy all the necessary requirements and do not result in any clashes with notions of scientific realism. 

\bibliographystyle{spphys}
\bibliography{eor_qm}

%\begin{acknowledgements}
%\end{acknowledgements}


\end{document}